\documentclass[prd, twocolumn, nofootinbib, floatfix]{revtex4-1}

\usepackage{amsmath}
\usepackage{graphicx}
\usepackage{dcolumn}
\usepackage{bm}
\usepackage{epsfig}
\usepackage{amssymb,latexsym,mathrsfs}
\usepackage{graphicx}
\usepackage{color}
\usepackage{hyperref}

\hypersetup{
    colorlinks=true,
    linkcolor=red,
    citecolor=blue,
} 

\newcommand{\be}{\begin{equation}}
\newcommand{\ee}{\end{equation}}
\newcommand{\bs}{\begin{split}} 
\newcommand{\bea}{\begin{eqnarray}}
\newcommand{\eea}{\end{eqnarray}}

\newcommand{\om}{\Omega_m}

\newcommand{\kmax}{k_{\rm max}}

\begin{document}

\title{Power Spectrum Precision for Redshift Space Distortions} 
\author{Eric V.\ Linder$^{1,2}$ \& Johan Samsing$^{1,3}$} 
\affiliation{$^1$Berkeley Center for Cosmological Physics \& Berkeley Lab, 
University of California, Berkeley, CA 94720, USA\\ 
$^2$Institute for the Early Universe WCU, Ewha Womans University, 
Seoul, Korea\\ 
$^3$Dark Cosmology Centre, Niels Bohr Institute, University of Copenhagen, 
Juliane Maries Vej 30, 2100 Copenhagen, Denmark}

\begin{abstract}
Redshift space distortions in galaxy clustering offer a promising 
technique for probing the growth rate of structure and testing dark 
energy properties and gravity.  We consider the issue of to what 
accuracy they need to be modeled in order not to unduly bias cosmological 
conclusions.  Fitting for nonlinear and redshift space corrections to the 
linear theory real space 
density power spectrum in bins in wavemode, we analyze both the effect of 
marginalizing over these corrections and of the bias due to not correcting 
them fully.  While naively subpercent accuracy is required to avoid bias in 
the fixed case, in the fitting approach the Kwan-Lewis-Linder reconstruction 
function for redshift space distortions is found to be accurately 
selfcalibrated with little degradation in dark energy and gravity parameter 
estimation for a next generation galaxy redshift survey such as BigBOSS. 
\end{abstract}

\date{\today} 

\maketitle

\section{Introduction} 

The pattern of galaxy clustering in three dimensions, and its evolution, 
encodes abundant information on the cosmological parameters affecting 
matter growth.  Ongoing and next generation spectroscopic galaxy surveys 
will vastly increase our measurements of this clustering, and our 
knowledge of cosmology if we can accurately interpret the results in terms 
of theory.  Measurements accrue an extra contribution to the redshift, and 
hence apparent position along the sight, from the galaxy peculiar velocities 
induced by the inhomogeneous density field; this gives rise to an anisotropy 
in the observed clustering known as redshift space distortions (RSD). 

These distortions carry information on the growth rate, as opposed to 
just the growth amplitude, and so are valuable for probing cosmology, as 
well as the gravitational strength driving the growth.  However, linear 
theory is insufficient for accurate relation of the redshift space galaxy 
power spectrum to the true (real space) matter density power spectrum, 
even on quite large scales, or wavenumbers $k>0.05\,h$/Mpc, where the vast 
majority of the statistical leverage lies 
\cite{percwhite,okujing,jennings1,jennings2,tang,kll}.  
Numerous corrections involving higher order perturbation theory have been 
employed \cite{scoccimarro,taruya,reidwhite,seljakmoment} 
that extend the validity but the region $k>0.1\,h$/Mpc is still 
problematic, especially for quantities involving the growth rate and 
the gravitational growth characterization.  For example, \cite{kll} 
demonstrates that these first principles approaches deliver results for 
the growth rate that are biased by several standard deviations when using 
modes out to $k=0.2\,h$/Mpc. 

Here we investigate a basic question: how accurately does one actually need 
to know the redshift space distortion mapping in order to extract the 
cosmological and gravitational parameter information without substantial 
bias or degradation?  Similar questions have been considered for weak 
gravitational lensing, for example, where one asks how well the nonlinear 
matter power spectrum needs to be known to estimate cosmology from the 
lensing shear power spectrum \cite{huttak,hearin}.  

In Section~\ref{sec:method} we introduce the correction, or reconstruction, 
function for the redshift space power spectrum and review the KLL \cite{kll} 
form for it.  Section~\ref{sec:bias} uses the Fisher bias method to 
compute both the individual parameter bias and joint confidence contour 
bias due to misestimated RSD, thus giving criteria for the accuracy to 
which the RSD effects must be known.  Adding fit parameters for uncertainties 
in the reconstruction function in Sec.~\ref{sec:marg}, we assess the 
impact of marginalizing over them on the cosmological parameters, in 
particular for tests of dark energy and gravity.  Section~\ref{sec:concl} 
summarizes the results and conclusions.

\section{Galaxy Power Spectrum} \label{sec:method}

\subsection{Redshift Space Power Spectrum} \label{sec:model} 

In real space the matter density power spectrum is expected to be 
isotropic, and the linear power spectrum grows in a scale independent 
manner through the growth factor $D(z)$, where $z$ is the redshift.  
The observed, redshift space galaxy power spectrum involves a transformation 
to redshift space due to the velocity effects, and a bias relation $b(z)$, 
usually taken to be scale independent, converting the dark matter overdensity 
to galaxy overdensity, and the effects of nonlinear structure formation. 
Each of these is modeled in various ways, with attendant uncertainties. 

We write the anisotropic redshift space galaxy power spectrum as 
\be 
P(k,\mu,z)=P^r(k,z)\,M(k,\mu,z)\,F(k,\mu,z)  \label{eq:pfactors} 
\ee 
where $P^r$ is the isotropic real space matter power spectrum, $M$ is an 
approximate model for redshift space distortions (including galaxy bias), 
and $F$ is the reconstruction function accounting for nonlinearities and 
more exact velocity effects. 

The linear mass power spectrum $P^r$ is given by 
a Boltzmann numerical code such as CAMB \cite{camb}.  It depends on 
the cosmological parameters through its shape as a function of wavenumber 
$k$ and through the growth factor $D(z)$ giving its amplitude evolution.  
Since we will correct the RSD modeling by the reconstruction function, we choose 
$M$ to be simply given by the linear theory prediction, the Kaiser approximation 
\cite{kaiser87}, 
\be 
M(k,\mu,z)=[b(z)+f(z)\,\mu^2]^2 \ , 
\ee 
where $b$ is the galaxy bias, $f=d\ln D/d\ln a$ the growth rate of 
density perturbations that in the linear regime grow as $\delta\sim D(a)$, 
where the scale factor $a=1/(1+z)$, and $\mu$ is the cosine of the angle made 
by the perturbation wavevector $\vec k$ with respect 
to the line of sight.  Beyond the linear regime, $b$ could be scale 
dependent, i.e.\ $b(k)$, but we will absorb that possibility into the 
reconstruction function.  The reconstruction function is fitted to N-body 
simulations by the analytic form of Kwan, Lewis, \& Linder (KLL; \cite{kll}), 
\be 
F(k,\mu,z)=\frac{A(k,z)}{1+B(k,z)k^2\mu^2}+C(k,z)k^2\mu^2 \ . \label{eq:fform} 
\ee 

This form has been found to reproduce accurately results of N-body 
simulations over a wide range of redshifts, and for halos of various 
masses as well as dark matter; see \cite{kll,julithesis} for details.  
Note that $A$, $B$, $C$ may be cosmology dependent, just as $f$ and $P^r$ 
are, and their universality is a subject of ongoing research, but here 
we treat them as independent parameters (as an analogy, recall how people 
treat coefficients within Halofit also as universal, though here we let the 
values of $A$, $B$, $C$ float; also see Sec.~\ref{sec:halofit}).  
The factor $A$ characterizes nonlinearity of 
the real space power spectrum, $B$ describes velocity effects such as 
damping from a Lorentzian velocity dispersion but also higher order multipole 
terms, while $C$ describes nonlinear enhancement for large $k\mu$ and 
possibly breaks the degeneracy in the two roles of $B$.

\subsection{Galaxy Clustering Information} 

The cosmological information inherent in the galaxy power spectrum can 
be estimated through the Fisher information matrix.  The full set of 
parameters $\{p_i\}$ includes the cosmological parameters, astrophysical 
parameters such as galaxy bias, and parameters for the reconstruction 
function.  Sensitivity 
to cosmology enters through the derivatives $\partial P/\partial p_i$ 
and the error covariance matrix for the redshift space galaxy power 
spectrum $P$. 

We follow the usual prescription \cite{fkp,seoeis03,stril} where the 
covariance 
matrix comes from sample variance (finite volume) and shot noise (finite 
resolution of the density field by sparse galaxies).  Taken together, the 
error can be thought of as depending on the number of modes that the galaxy 
redshift survey samples.  Treated as Poisson sampling of the density field, 
the statistical error is 
\be 
\sigma_P=P+n^{-1} 
\ee 
from these two effects.  The number of Fourier modes is the volume of 
a Fourier cell times the number of cells, 
\be 
N_{\rm modes}=2\pi k^2 dk\,d\mu \times V_{\rm survey}/(2\pi)^3 \ . 
\ee 
Therefore the error covariance matrix $C$ is 
\be 
C=P^2\,\left(\frac{1+nP}{nP}\right)^2 \, \frac{8\pi^2}{V_{\rm survey}k^2 dkd\mu} \ . 
\ee 

Since the Fisher information matrix is constructed from $C^{-1}$ multiplied 
by the sensitivity derivatives $\partial P/\partial p_i$, we can use the 
$P^2$ factor to convert the derivatives to involve $\ln P$, which will 
be useful in treating the multiplicative factors in Eq.~(\ref{eq:pfactors}). 
In summary, the Fisher matrix is 
\be 
F_{ij}=\sum_{z}\sum_{\mu}\sum_{k}\,\frac{\partial\ln P}{\partial p_i} 
\frac{\partial\ln P}{\partial p_j}\, V_{\rm eff}(k,\mu,z) 
\frac{k^2\Delta k\,\Delta\mu}{8\pi^2} \ , 
\ee 
where the survey volume is reduced by the shot noise to a $z$, $k$, and 
$\mu$ dependent effective volume 
\be 
V_{\rm eff}(k,\mu,z)=V_{\rm survey}(z)\, 
\left[\frac{n(z)P(k,\mu,z)}{n(z)P(k,\mu,z)+1}\right]^2 \ . \label{eq:veff} 
\ee 
When the galaxies densely sample the underlying field, the effective 
volume approaches the survey volume, otherwise modes are lost, diluting 
the effective volume due to increased noise. 

Note that the logarithmic derivatives can be written as 
\bea 
\frac{\partial\ln P}{\partial p_i}\,\frac{\partial\ln P}{\partial p_j}&=& 
\frac{(\partial\ln P^r+\partial\ln M+\partial\ln F)}{\partial p_i}\notag\\  
&\qquad&\times\frac{(\partial\ln P^r+\partial\ln M+\partial\ln F)}{\partial p_j}\, 
\eea 
so only the $\partial\ln F$ term depends on $A$, $B$, and $C$. 

The reconstruction function derivatives are 
\bea 
\frac{\partial F}{\partial A}&=&\frac{1}{1+Bk^2\mu^2} \\ 
\frac{\partial F}{\partial B}&=&\frac{-Ak^2\mu^2}{(1+Bk^2\mu^2)^2} \\ 
\frac{\partial F}{\partial C}&=&k^2\mu^2 \ ,  
\eea 
and are otherwise taken not to depend on cosmology.  This is because 
we use $A$, $B$, $C$ purely as fiducial values, and investigate how 
their variation (from astrophysics or cosmology) impacts the cosmological 
parameter estimation.  

\begin{figure}[htbp!]
\includegraphics[width=\columnwidth]{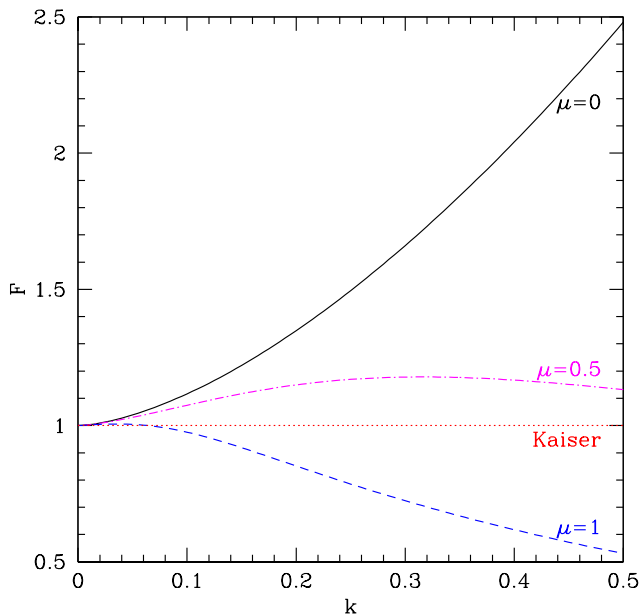}
\caption{The redshift space distortion reconstruction function $F(k,\mu)$ 
is plotted for the fiducial expressions for $A$, $B$, $C$ for three values 
of angle $\mu$. 
} 
\label{fig:fplot} 
\end{figure}

Our fiducial case attempts to match $F$ to the simulation 
results in \cite{kll}, with estimated 
\bea 
A(k)&=&1+\left(\frac{k}{0.39\,h/{\rm Mpc}}\right)^{1.58} \label{eq:aform}\\ 
B(k)&=&20\,({\rm Mpc}/h)^2 \label{eq:bform}\\ 
C(k)&=&8\,e^{-k/(0.176 h/{\rm Mpc})}\,({\rm Mpc}/h)^2 \ . \label{eq:cform} 
\eea 
The resulting redshift space distortion reconstruction function of 
Eq.~(\ref{eq:fform}) is shown in Fig.~\ref{fig:fplot}.  
We emphasize that these are merely the fiducials; we allow 
the values of $A$, $B$, $C$ to float freely in bins of wavenumber.  
This provides a model independent variation of the power spectrum (within 
the reconstruction form) and we can then investigate the influence of such 
variations on the cosmological parameter estimation.  Conversely, the 
question can be phrased as ``what is the accuracy 
required on knowledge of the galaxy power spectrum in order to deduce the 
cosmology with confidence?''  

We later contrast this fiducial with fiducial $(A,B,C)=(1,0,0)$, i.e.\ 
assuming that perturbation theory (for example linear theory in the Kaiser 
case, although $F$ also works with higher order perturbation theory 
\cite{kll}) fully captures RSD effects in the model $M$.

The analysis is carried out in the next sections in two ways: in 
Sec.~\ref{sec:bias} we compute the effect that a given level of unrecognized 
power spectrum deviation in some $k$ bin, i.e.\ a systematic error in 
modeling, has in biasing the cosmological conclusions, and in 
Sec.~\ref{sec:marg} we recognize the existence of systematic uncertainties 
and treat them by marginalizing 
over the $A$, $B$, $C$ values for each $k$ bin.

\subsection{Survey Characteristics and Parameters} \label{sec:survey} 

For the galaxy redshift survey data we consider a next generation 
spectroscopic survey of the quality proposed for BigBOSS \cite{bigboss}, 
covering 14000 deg$^2$ from $z=0.1-1.8$, with a galaxy number density $n$ of 
approximately $3\times 10^{-4}\,h^3\,{\rm Mpc}^{-3}$.  For the exact 
distribution adopted see Table~\ref{tab:nz}.  There are actually two 
populations of galaxies: luminous red galaxies (LRG) and emission line 
galaxies (ELG), each with their own galaxy bias value.  These galaxy 
biases are taken as free parameters to be marginalized over, for each 
redshift bin of width $0.1$.  Their fiducials are $b(z)=b_0 D(z=0)/D(z)$ 
with $b^{\rm ELG}_0=0.8$ and $b^{\rm LRG}_0=1.6$, which provide good fits 
to observations.  Galaxy populations 
with different biases can help reduce sample variance \cite{selmc}, with 
the Fisher matrix involving a sum over populations, i.e. 
\be 
\sum_{XY} 
\frac{\partial\ln P_X}{\partial p_i}\frac{\partial\ln P_Y}{\partial p_j} 
\,\left[\frac{n_X P_X}{n_XP_X+1}\right] \left[\frac{n_Y P_Y}{n_YP_Y+1}\right] 
\ . 
\ee 
Note that for multiple populations the factor $V_{\rm eff}$ in 
Eq.~(\ref{eq:veff}) involves the shot noise, i.e.\ $nP$, of each population.

\begin{table}[!htb]
\begin{tabular}{ccc}
$z$ & $\ n_{\rm ELG}\ $ & $\ n_{\rm LRG}\ $\\ 
$\ $0.15$\ $ & 22.6 & 30.1\\ 
0.25 & 8.45 & 3.04\\ 
0.35 & 4.02 & 3.07\\ 
0.45 & 2.65 & 3.09\\ 
0.55 & 2.99 & 3.10\\ 
0.65 & 3.99 & 3.11\\ 
0.75 & 5.15 & 3.12\\ 
0.85 & 5.36 & 1.89\\ 
0.95 & 5.02 & 0.33\\ 
1.05 & 4.80 & 0.04\\ 
1.15 & 4.49 & --\\ 
1.25 & 4.04 & --\\ 
1.35 & 3.02 & --\\ 
1.45 & 2.00 & --\\ 
1.55 & 1.15 & --\\ 
1.65 & 0.43 & --\\ 
1.75 & 0.12 & --\\ 
\end{tabular}
\caption{Spectroscopic survey number densities adopted for emission line 
galaxies and luminous red galaxies, in units of $10^{-4}\,h^3/{\rm Mpc}^3$, 
for each redshift shell. 
}
\label{tab:nz}
\end{table}

For the cosmological parameters we use the physical baryon density 
$\Omega_b h^2$ and physical cold dark matter density $\Omega_c h^2$, 
reduced Hubble constant $h$, scalar perturbation tilt $n_s$ and 
amplitude $A_s$, dark energy equation of state parameters $w_0$ 
and $w_a$, and gravitational growth index $\gamma$.  The gravitational 
growth index gives an accurate description of the growth rate for both 
general relativity and a range of modified gravity models, and looking 
for deviations from its general relativistic value of $\gamma=0.55$ acts 
as a test of gravity \cite{groexp,lincahn}.  The growth index is 
treated as an independent parameter (not a function of $w_0$, $w_a$) and 
determines the growth factor at scale factor $a=1/(1+z)$, 
\be 
D(a)=e^{\int_0^a (da'/a')\,\om(a')^\gamma} \ , 
\ee 
that in this ansatz is used to convert the linear power spectrum 
delivered by CAMB at $z=0$ to another redshift $z$, to account for 
the effects of modified gravitational growth.  Note that the growth 
rate $f=\om(a)^\gamma$, and redshift space distortions were highlighted 
as a test of gravity in \cite{linrsd}. 

For the central question of RSD uncertainties we employ up to 12 
free parameters, taking $A$, $B$, $C$ with independent values in each 
bin of width $0.1$ in wavenumber above $k=0.1\,h$/Mpc out to some $\kmax$.  
This corresponds 
to uncertainties $\Delta P_k$.  For the current work we follow 
\cite{huttak,hearin} and consider the uncertainties only as a function 
of wavenumber, not redshift, except in Sec.~\ref{sec:reddep}; 
we also take the KLL form to be accurate while allowing freedom in the 
parameters $A$, $B$, $C$.  In summary we fit 
for 8 cosmological parameters and up to 39 systematics parameters.

\section{Fisher Bias} \label{sec:bias} 

The first question we are interested in answering is what is the 
sensitivity of the cosmological parameter determination to errors in 
modeling RSD.  One might have $M$ or $F$ wrong, but if this does not 
mimic a change in cosmology then no harm is done.  The Fisher bias 
formalism (see, e.g., \cite{fisbias1,fisbias2}) propagates misestimation 
of the 
observable or theoretical prediction, in this case the redshift space 
power spectrum, into biases on the fit parameters.  Specifically, we 
consider the effect of errors in the $k$ bin values of $A$, $B$, $C$. 

The Fisher bias on a parameter $p_i$ from misestimating parameter $p_a$ is 
\be 
\delta p_i=\delta p_a \, \sum_{j} (F^{\rm sub})^{-1}{}_{ij}\, 
(F^{\rm full})_{ja} \ , 
\ee 
where $\delta p_a=p_a({\rm true})-p_a({\rm fiducial})$.  The superscript 
``sub'' denotes the Fisher submatrix without entries for the parameters 
whose misestimation we are studying.  (For the specific case here, the 
submatrix will be $35\times 35$ for the cosmology and galaxy bias parameters, 
and the full matrix adds the reconstruction parameters one at a time.  
We later consider all the reconstruction parameters at once.)  
By evaluating the ratio $dp_i/dp_a$ for 
$a=A,B,C$ we can assess the sensitivity of the parameter estimation to the 
RSD modeling.  To take a weak lensing example, \cite{berkwl} found that 
a particular form of matter power spectrum distortion with amplitude 
$A_{NL}$ at high $k$ distorted estimation of $w_a$ derived from shear power 
spectrum measurement by a leverage factor of 18: a misestimation of 10\% 
in $A_{NL}$ yielded a $1.8\sigma$ bias in $w_a$. 

The bias $\delta p_i$ can be compared to the statistical uncertainty 
$\sigma(p_i)$ on the parameter, either directly or through the risk 
statistic 
\be 
R_i\equiv \sqrt{\sigma^2(p_i)+\delta p_i^2} \ . 
\ee 
Treating the bias as a systematic error in this way, to restrict the 
degradation in the statistical error to less than 20\%, say, requires 
$\delta p_i/\sigma(p_i)<0.66$, thus putting a constraint on the allowable 
modeling error $\delta p_a$ etc.  We examine two, converse statistics: 
the cosmological degradation caused by a certain fractional misestimation 
of the reconstruction parameters $\delta p_a/p_a$, and the requirement 
on the reconstruction parameter to bound the cosmological parameter bias 
to less than a given factor of the statistical uncertainty, 
$\delta p_i/\sigma_i$.  These are respectively 
\bea 
\frac{R_i}{\sigma_i}&=&\sqrt{1+\left(\frac{\delta p_i}{\delta p_a}\frac{\delta p_a}{p_a} \frac{p_a}{\sigma_i}\right)^2} \\ 
\frac{\delta p_a}{p_a}&=&\left(\frac{\delta p_i}{\delta p_a}\right)^{-1} 
\frac{\delta p_i}{\sigma_i}\frac{\sigma_i}{p_a} \ . 
\eea 

\begin{figure}[htbp!] 
\includegraphics[width=\columnwidth]{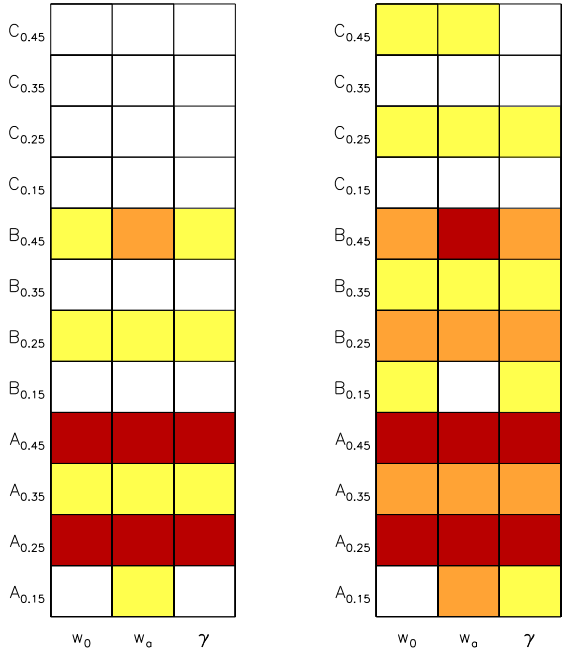} 
\caption{[Left panel] The ratio of the root mean squared error, or risk, to 
the statistical uncertainty, $R_i/\sigma_i$, is plotted for each dark 
cosmological parameter in the case of a 1\% deviation in a reconstruction 
parameter.  [Right panel] The fractional requirement on each reconstruction 
parameter $\delta p_a/p_a$ needed to ensure bias less than $1\sigma$, i.e.\ 
$\delta p_i/\sigma_i<1$ is plotted.  Dark red indicates danger (high risk 
or tight requirement), with lighter colors showing reduced impact.  For 
the left panel the color scale is $R_i/\sigma_i>2$ (dark red), 1.4--2 
(medium orange), 1.05--1.4 (light yellow), and 1--1.05 (white).  The right 
panel has $|\delta p_a/p_a|<0.01$ (dark red), 0.01--0.05 (medium orange), 
0.05--0.2 (light yellow), $>0.2$ (white).  Here $\kmax=0.5\,h$/Mpc. 
}
\label{fig:degr}
\end{figure}

Figure~\ref{fig:degr}, left panel, shows the matrix of 
degradations $R_i/\sigma_i$ for fixed $\delta p_a/p_a=0.01$ (i.e.\ 1\% 
uncertainty on the reconstruction parameters), where the columns are the 
dark cosmological parameters and the rows are the reconstruction parameters. 
The right panel gives a similar matrix of the reconstruction 
requirements $\delta p_a/p_a$ for fixed $\delta p_i/\sigma_i=1$ (which 
corresponds to $R_i/\sigma_i=1.41$).  One can scale the results for 
different fixed values according to the above equations.  
The stripe structure arises because the bias from $A_{0.45}$, where the 
subscript indicates the center of the $k$ bin, is of 
opposite sign from that of $A_{0.25}$, and $A_{0.35}$ lies in between 
near null effect, and similar for $B$ and $C$. 

The degradations in determination of the dark parameters $w_0$, $w_a$, 
$\gamma$ are less than 22\% for 1\% shifts in the reconstruction parameters 
in all cases except $A_{0.25}$, $A_{0.45}$, and $B_{0.45}$.  
For $A_{0.25}$ and $A_{0.45}$ the risk error on $w_a$ can exceed the 
statistical uncertainty by a factor 3.  For the $B$ parameters the worst 
case is degradation by 1.5.  These results offer indications of what 
physics must be most accurately understood, i.e.\ the nonlinearity from 
$A$ and, somewhat less critically, the velocity effects from $B$. 

In the converse analysis of what accuracy is required on the reconstruction 
parameters to ensure that a bias is restricted to below $1\sigma$, we 
find that 5\% accuracy is sufficient for all parameters except for the 
$A_k$, plus $B_{0.25}$ and $B_{0.45}$.  Knowledge of $A_{0.25}$ and 
$A_{0.45}$ are needed to 0.3\%, $B_{0.45}$ to 0.9\%, $B_{0.25}$ to 1.4\%, 
$A_{0.35}$ to 1.5\%, and $A_{0.15}$ to 3.1\%.

While these approaches give indications of sensitivity, they 
treat the cosmological parameters one by one while a power 
spectrum misestimation will generally impact several at once.   This can 
either tighten or loosen overall requirements, depending on the 
covariances.  To take this 
into account we use the $\Delta\chi^2$ method \cite{dodelsonbias,shapiro}. 
This describes the fuller impact of biasing cosmology through quantifying 
how far from the fiducial the best fit cosmology is shifted relative to the 
confidence contour, taking into account degeneracies between the 
reconstruction and cosmological parameters.  This measure is given by 
\be 
\Delta\chi^2=\sum_{ij} \delta p_i\,F^{({\rm red})}_{ij}\,\delta p_j \ , 
\ee 
where the sum runs only over the reduced parameter set whose bias we 
are interested in, e.g.\ $w_0$ and $w_a$ for a 2D $w_0$--$w_a$ joint 
likelihood 
contour plot.  The reduced Fisher matrix $F^{({\rm red})}$ is marginalized 
over all other cosmological and galaxy bias parameters (the reconstruction 
parameters have already been taken into account in obtaining $\delta w_0$ 
etc.).  The bias $\Delta\chi^2$ accounts for the property that biases in, 
say, the direction of the narrow axis of the Fisher ellipse are more 
detrimental than those along the degeneracy direction. 

\begin{figure}[htbp!]
\includegraphics[width=\columnwidth]{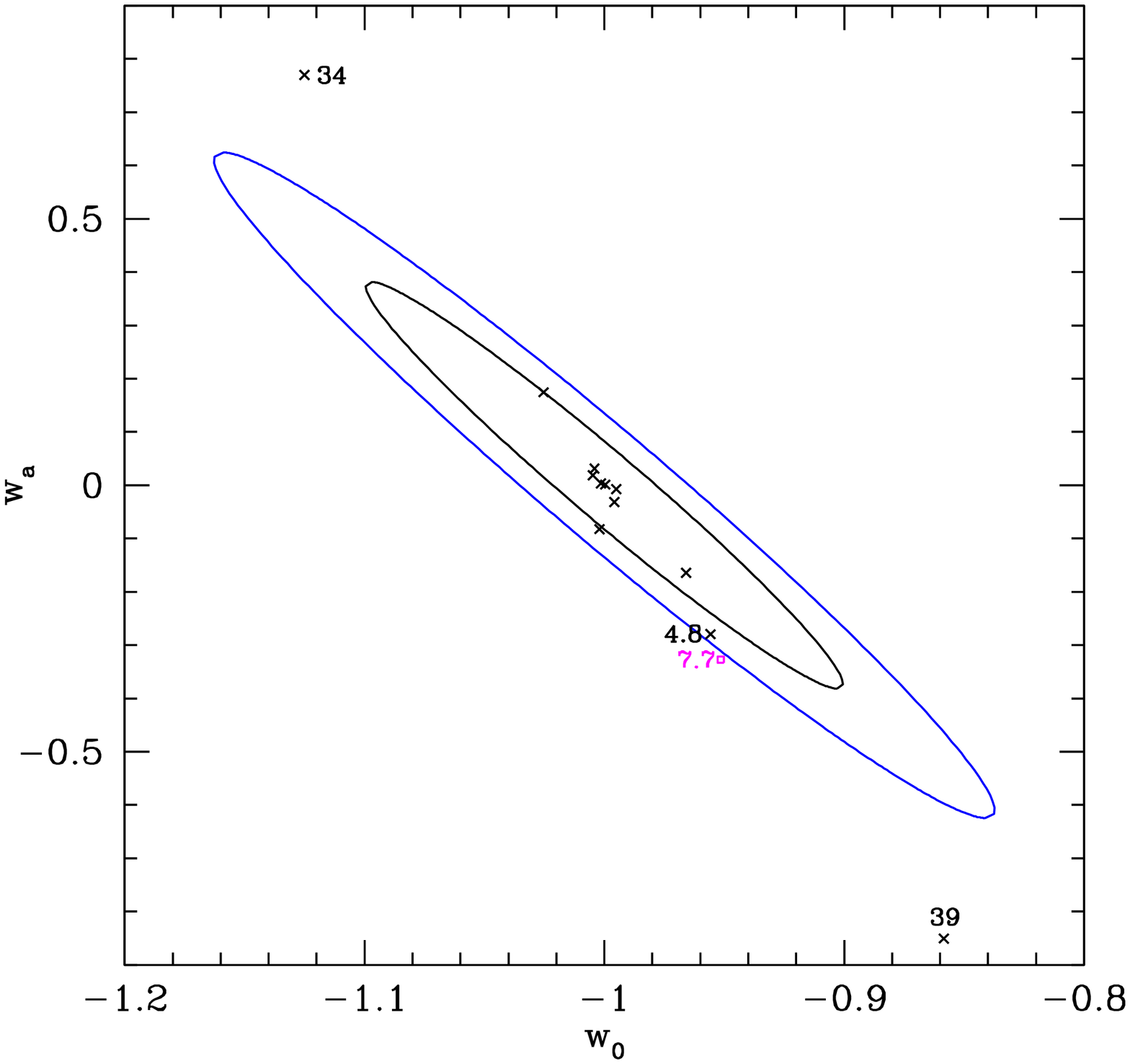} \\ 
\includegraphics[width=\columnwidth]{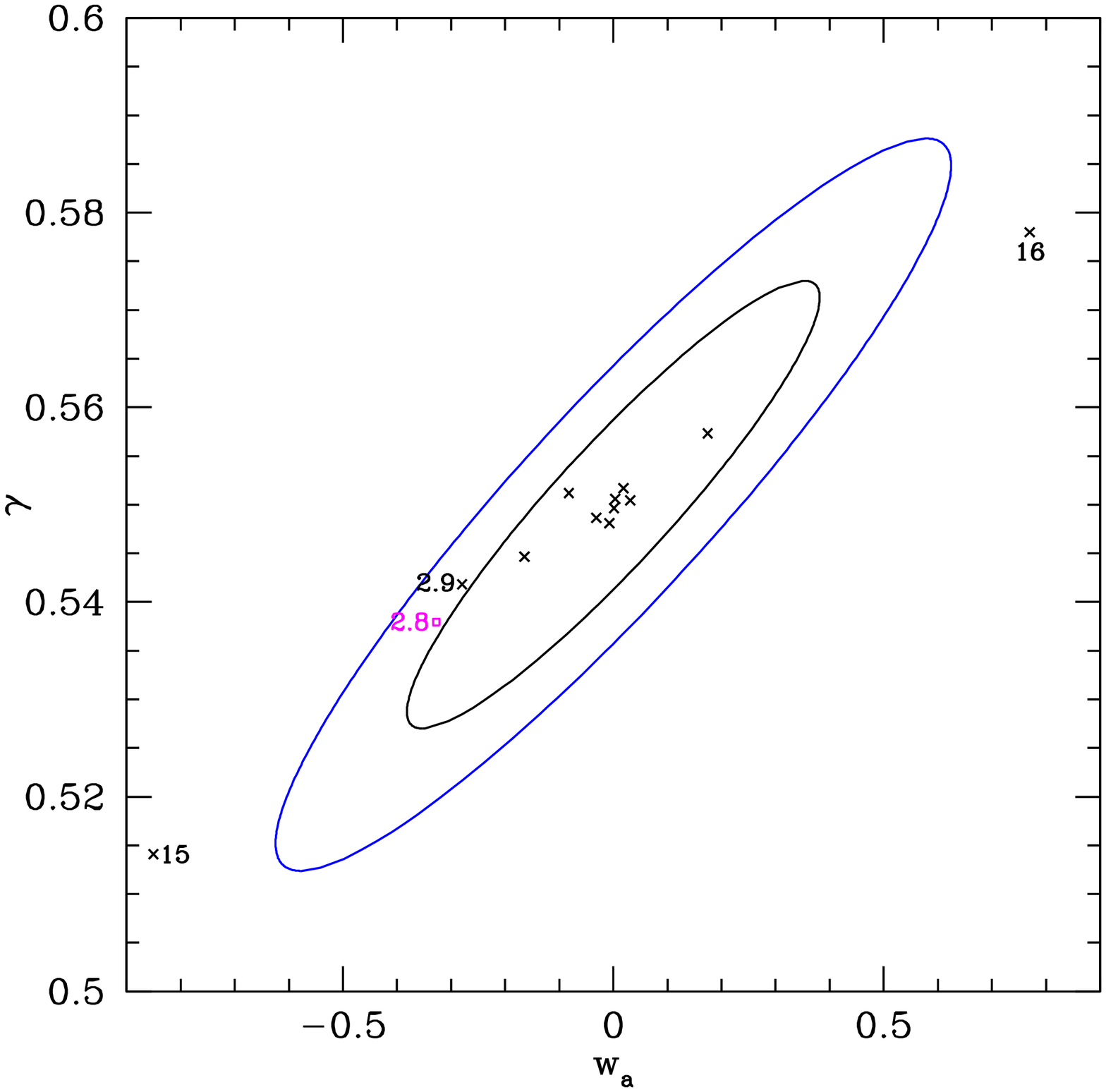} 
\caption{The biases in the $w_0$--$w_a$ and $w_a$--$\gamma$ planes 
due to 1\% misestimation in the 12 reconstruction parameters, one by one, 
are shown by x's (along with the $\Delta\chi^2$ if larger than 2.8).  
The contours give the joint 2D $1\sigma$ and $2\sigma$ confidence levels 
on the cosmological parameters when the reconstruction parameters take 
their fiducial (``true'') values.  Magenta squares show the biases when 
varying all bins of $A$ simultaneously; such smooth variations are much 
less damaging, e.g.\ reducing the individual $\Delta\chi^2=39$ and 34 
biases to a joint $\Delta\chi^2=7.7$ offset (or the 16 and 15 in the 
$w_a$--$\gamma$ panel to 2.8). 
}
\label{fig:chi2}
\end{figure}

Figure~\ref{fig:chi2} illustrates the 2D bias induced in the dark energy 
and growth parameters, here for a 1\% misestimation in the reconstruction 
parameters one by one.  Most such reconstruction parameter errors do not 
significantly affect the joint parameter likelihood.  In the $w_0$--$w_a$ 
plane, none of the $C$ parameters and three of the $B$ parameters do not 
bias the best fit outside the $1\sigma$ contour, and 
$B_{0.45}$ remains within the $2\sigma$ contour.  Only errors on the 
$A_{0.25}$ and $A_{0.45}$ parameters are particularly damaging, causing a 
bias of up to $\Delta\chi^2=39$ (approximately at the $6\sigma$ level).  
The covariance between the shifts in $w_0$ and $w_a$ is crucial; if the 
same bias in $w_a$ and an even larger bias in $w_0$ occurred such that 
the joint shift lay along the degeneracy axis, then the 2D bias would be 
scarcely outside the $2\sigma$ uncertainty contour.  
For the $w_a$--$\gamma$ plane, the biases are less severe, with only 
$A_{0.25}$ and $A_{0.45}$ causing more than a $2\sigma$ joint bias, at 
$\Delta\chi^2\approx 16$. 

Treating the reconstruction errors one by one effectively takes a 
localized bump in the reconstruction function.  A smooth variation would 
instead affect several of the reconstruction parameters 
at once; this has a different effect on the cosmological parameter bias. 
As an example, we simultaneously vary all four $A$ parameters by 1\%.  Since 
$A_{0.25}$ and $A_{0.45}$ have nearly opposite effects this reduces 
substantially the $\Delta\chi^2$ due to varying just one of them, e.g.\ from 
39 to 7.7.  The 2D bias due to such smooth variation is shown in the figures 
by the magenta squares.

While we have thus far been model independent in taking $A$, $B$, $C$ 
to be independent from one $k$ bin to the next, we can now consider 
the difference between two fiducial models for the overall reconstruction 
function $F$.  This then includes the effects of variations at all $k$'s 
simultaneously, and allows a study of bias as a function of $\kmax$.  
As we consider 
each successive bin at higher $k$, we increase the number of modes, 
reducing the statistical uncertainty, but also often increase the deviation 
in the power spectra, increasing the bias in the cosmological parameters 
if we assume the wrong fiducial as the truth.  The truth is taken to be 
$F$ as given by Eqs.~(\ref{eq:aform})--(\ref{eq:cform}) in 
Eq.~(\ref{eq:fform}), while the incorrect assumption is pure linear theory, 
i.e.\ simply the Kaiser form for redshift space distortions, equivalent to 
$A=1$, $B=C=0$. 

This misestimation of the redshift space galaxy power spectrum causes a 
bias in cosmological parameters of 
\bea 
\!\!\delta p_i&=&\left (F^{\rm sub}\right)^{-1}_{ij} \sum_z \sum_\mu \sum_k \, 
\frac{P(A,B,C)-P(1,0,0)}{P(A,B,C)} \notag\\ 
&\qquad&\times\frac{\partial\ln P}{\partial p_j}\, V_{\rm eff}(k,\mu,z) 
\frac{k^2\Delta k\,\Delta\mu}{8\pi^2} \ . 
\eea 
The systematic biases tend to worsen with increasing $\kmax$, reaching 
$1.4$ in $w_0$, $-8$ in $w_a$, and 0.2 in $\gamma$ for $\kmax=0.5\,h$/Mpc, 
and are much larger than the statistical uncertainties for all 
$\kmax$.  This is not surprising since $F_{\rm Kaiser}$ can deviate 
by a factor 2 from the KLL form.  Thus, neglecting the uncertainties in 
the reconstruction parameters is not a viable option: we must take them 
into account.

\section{Marginalization and Selfcalibration} \label{sec:marg} 

As an alternative to requiring the power spectrum to subpercent accuracy 
and computing the bias from misestimated reconstruction 
parameters, we can fit for those parameters and calculate the increased 
uncertainty in cosmological parameters due to marginalization over the 
extra inputs.  The basic question is how well the model needs to be 
known for precision determination of cosmology with RSD.  This is similar 
to what \cite{huttak,hearin} did for matter power 
spectrum uncertainties applied to weak lensing cosmology.  They used 
fractional power spectrum uncertainties in wavenumber bins, assumed 
constant with redshift, and applied some level of priors.

\subsection{Global Fit} \label{sec:global} 

Now our quantities $A_k$, $B_k$, $C_k$ in each wavenumber bin become 
fit parameters.  Again, we can study the effects as we extend $\kmax$, 
using more bins and hence more parameters.  Including these parameters 
means that we will not be biased any more with respect to the fiducial, 
but the enlarged parameter space will lead to some level of degradation 
of the statistical uncertainties, relative to fixing the reconstruction 
parameters, at the same $\kmax$.  

Table~\ref{tab:sigkmax} shows the effect of extending the data to 
higher $\kmax$, while simultaneously allowing for the additional 
reconstruction parameters in each bin.  Despite the additional degrees 
of freedom in the fit, the cosmological parameter estimation sharpens 
with increasing $\kmax$.  As long as the {\it form\/} of the reconstruction 
function holds, we obtain an accurate and unbiased cosmology even allowing 
for fitting variation in the amplitudes of $A$, $B$, $C$ in each $k$ bin.  
This is an extremely promising initial result for use of the reconstruction.

\begin{table*}[!htb]
\begin{tabular}{l|ccccccccc}
&$\Omega_b h^2$ &$\Omega_c h^2$ &$h$&$n_s$&$10^9 A_s$&$w_0$&$w_a$&$\gamma$& 
$\Omega_m$\\ 
\hline 
Fiducial& $\ $0.0226& 0.112& 0.7& 0.96& 2.47& -0.99& 0& 0.55& 0.275\\ 
$\sigma(\kmax=0.1)$\ & $\ $ 0.00524 &  0.0189 &  0.0542 &  0.0524 &  0.538 & 
  0.599 &  2.23 &  0.177 & 0.0302\\ 
$\sigma(\kmax=0.2)$\quad & $\ $0.00284 &  0.0102 &  0.0284 &  0.0288 &  0.325 & 0.197 & 0.779 & 0.0519 & 0.0159\\ 
$\sigma(\kmax=0.3)$\ & $\ $0.00219 & 0.00760 &  0.0219 &  0.0198 &  0.248 & 
  0.112 &  0.478 &  0.0272 & 0.0122\\ 
$\sigma(\kmax=0.4)$\ & $\ $ 0.00148 &  0.00508 &  0.0150 &  0.0130 &  0.170 &
  0.0824 &  0.347 &  0.0193 & 0.00834\\ 
$\sigma(\kmax=0.5)$\ &  $\ $0.00141 & 0.00478 & 0.0142 & 0.0119 & 0.158 & 0.0713 & 0.306 & 0.0163& 0.00794\\  
\end{tabular}
\caption{$1\sigma$ constraints from future galaxy power spectrum data on 
cosmological parameters, marginalized over galaxy bias and redshift space 
distortion reconstruction.  Note $\om$ is a derived parameter; $\kmax$ is 
in units of $h$/Mpc.  Despite the addition of more fit parameters when 
increasing $\kmax$, the cosmological parameters can be better determined. 
}
\label{tab:sigkmax}
\end{table*}

\begin{figure}[htbp!]
\includegraphics[width=\columnwidth]{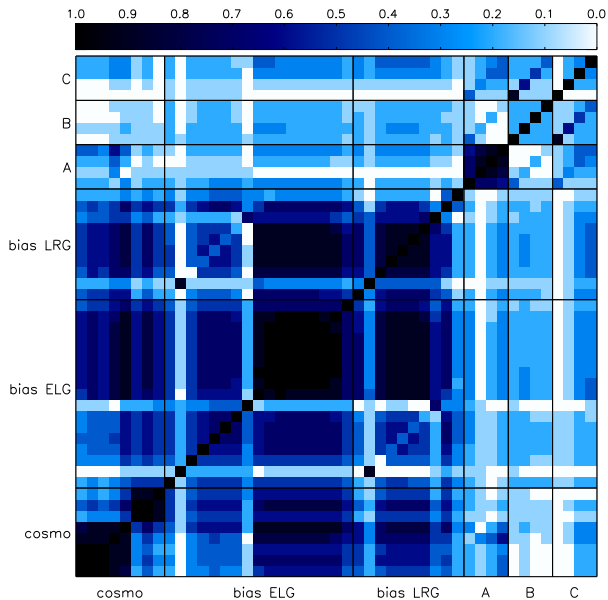}
\caption{Correlation matrix of the 47 parameters for $\kmax=0.5\,h$/Mpc is 
shown with color shading reflecting the absolute value of the correlation 
coefficient $r_{ij}=C_{ij}/\sqrt{C_{ii} C_{jj}}$.  The correlation matrix is 
mostly block diagonal and the cosmological parameters are not strongly 
correlated with the reconstruction (or galaxy bias) parameters, so 
marginalization does not badly degrade cosmological parameter estimation. 
}
\label{fig:corrmat}
\end{figure}

To better understand why the added fit parameters do not cause an overall 
degradation, we look at the correlation matrix of the 47 parameters in 
Fig.~\ref{fig:corrmat}.  The block of parameters 36--47, representing the 
reconstruction parameters, is not highly correlated with other parameters: 
correlation coefficients are below 0.58 (0.38 for parameters other than 
$n_s$).  
(Even within the block, only $B_{0.15}$ and $C_{0.15}$, other than 
between the $A_k$, have a correlation coefficient exceeding 0.8, 
reaching 0.90; this is expected 
since for a low $k$ expansion both $B$ and $C$ contribute as $\mu^2$.)  
This means that the change in power spectrum shape due to adjusting the 
amplitudes of these parameters in $F$ is not degenerate with a change due 
to $w_0$ or other such parameters.  That is, the influence of these 
parameters have different $k$ and $\mu$ dependences than those of 
cosmological parameters and so we find they can be separately fit.

Moreover, the reconstruction parameters are selfcalibrated by the data 
to good precision.  All are determined to better than 3\% (except 
$C_{0.15}$, to 8\%) and most to subpercent level.  These propagate into 
determination of the power spectrum to the subpercent level for variation 
of each one individually by $1\sigma$, except for the extreme cases of 
$\mu=1$ and $B_{0.15}$ 
($C_{0.15}$) which gives 1.1\% (1.2\%) uncertainty.  Most combinations, 
however, give subpercent precision.  Adding all their 
uncertainties in the most unfavorable way generates an extreme of 2.6\% 
power spectrum uncertainty.  
Thus unlike the weak lensing probe analyzed by \cite{huttak,hearin}, 
redshift space distortions do not require any priors to be placed on the 
power spectrum 
parameters (assuming that the KLL reconstruction form is valid). 

Remarkably, in addition 
to selfcalibration, the additional fit parameters have little impact on 
the cosmological parameter estimation, enlarging the uncertainties by 
only 9\%, 22\%, and 7\% on $w_0$, $w_a$, and $\gamma$.  And of course 
including the extra parameters removes any cosmology bias as suffered 
in the previous section (modulo model validity).  
Regarding the 12 extra reconstruction parameters, from Fig.~\ref{fig:fplot} 
we see that $F$ is smooth in $k$ so taking bins of width 0.1 in $k$ is 
reasonable.  For completeness, for bins of width 0.02 (and hence 60 extra 
parameters) we find that cosmological parameter estimation is mildly 
degraded, with uncertainties on $w_0$, $w_a$, $\gamma$ increasing relative 
to 0.1 width by 16\%, 33\%, 17\%.

\subsection{Maximum Wavenumber} \label{sec:kmax} 

\begin{figure}[htbp!]
\includegraphics[width=\columnwidth]{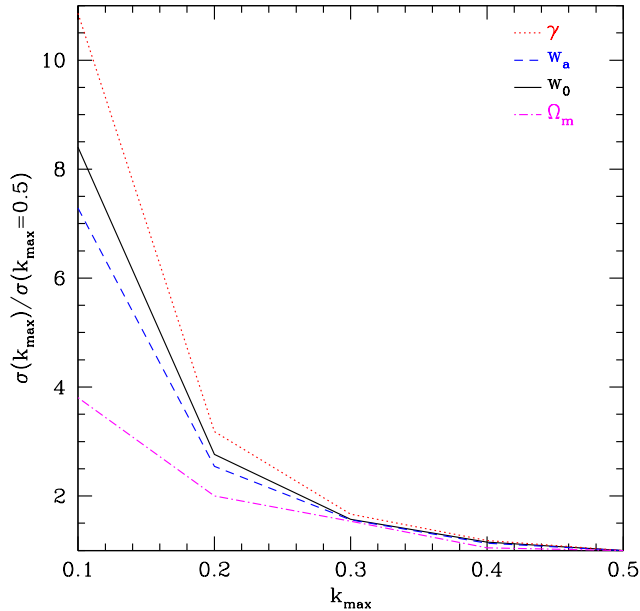}
\caption{Extending $\kmax$ to values above $0.1\,h$/Mpc breaks 
degeneracies, leading to improvements in cosmological parameter estimation 
as shown here, even given the addition of reconstruction parameters to 
marginalize over.  Reconstruction to $\kmax=0.4-0.5\,h$/Mpc is sufficient 
to plateau the cosmology estimation precision. 
} 
\label{fig:kmaxratio} 
\end{figure}

Let us examine the dependence of the results on the maximum wavenumber 
$\kmax$ used.  
Note that for the $\kmax=0.1$ case, the cosmology parameters are not 
determined particularly well even though no reconstruction parameters 
are used for $k\le0.1\,h$/Mpc.  This is due to strong covariance with 
the 27 galaxy bias parameters.  Once beyond the linear regime, this 
degeneracy is broken and the correlation coefficients drop, greatly 
improving the cosmological parameter determination (e.g.\ by factors of 
2.9--3.4 on the dark parameters, for $\kmax=0.2$ relative to $\kmax=0.1$).  
This continues for higher $\kmax$, 
despite the addition of further reconstruction parameters, but gradually 
saturates.  For example, while relative to the $\kmax=0.5$ case the 
uncertainties on $w_0$, $w_a$, or $\gamma$ at $\kmax=0.2$ are greater by 
a factor $\sim3$, at $\kmax=0.3$ the factor is 1.6, and at $\kmax=0.4$ 
is 1.15, as seen in Fig.~\ref{fig:kmaxratio}.  Thus, having an accurate 
reconstruction form out to $\kmax\approx 0.4-0.5$ is sufficient for 
robust cosmological parameter estimation, while selfcalibration obviates 
the need for any prior knowledge of the values of the reconstruction 
parameters.

Binning such as we use is model independent and closest to the weak 
lensing work.  
This model independence is important since in the absence of a large suite of 
simulations we may have no particular confidence in parametrizations such 
as those in Eqs.~(\ref{eq:aform}-\ref{eq:cform}).  Recall that those 
equations merely give the 
fiducial values in each $k$ bin, and then we allow the bin values to 
float freely and marginalize over them.  Given simulations we might be 
able to adopt specific forms and fit for a reduced set of parameters, 
e.g.\ the coefficients in those equations.

\subsection{Redshift Dependence} \label{sec:reddep} 

To give a first indication of whether adding redshift dependence to 
$A$, $B$, $C$ affects the conclusions we include a variation of the 
characteristic wavenumber scale entering in the nonlinearity amplitude 
$A$ in Eq.~(\ref{eq:aform}), i.e.\ the $0.39\,h$/Mpc, writing this as 
\be 
k_\star(z)=0.39\,(1+z)^{\alpha/1.58}\,h/{\rm Mpc} \ , 
\ee 
and adding this evolution parameter $\alpha$ to the fit.  The simulation 
results in \cite{julithesis} indicate that relative to $A$, the parameters 
$B$ and $C$ have negligible additional redshift dependence.  Therefore we 
scale $B$ and $C$ by the same factor as $A$, i.e.\ 
\bea 
B(k,z)&=&B(k,0)\,A(k,z)/A(k,0) \\ 
&=&B(k,0)\,\frac{1+[A(k,0)-1](1+z)^{-\alpha}}{A(k,0)} \ , 
\eea 
and the same for $C$. 

\begin{figure}[htbp!]
\includegraphics[width=\columnwidth]{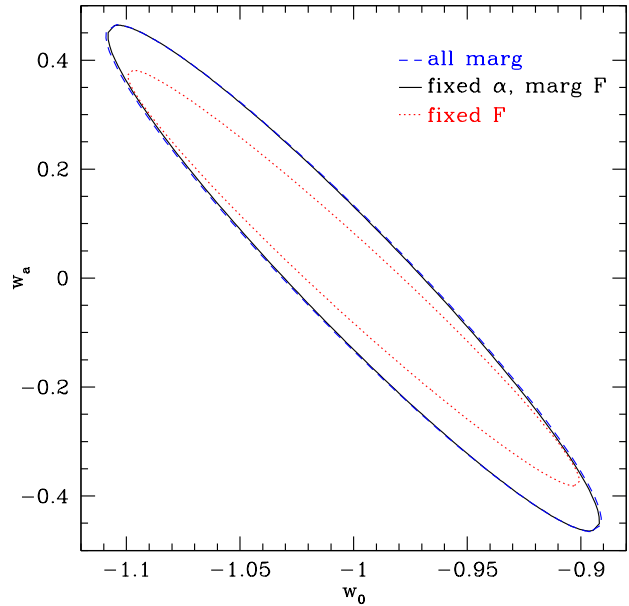}\\ 
\includegraphics[width=\columnwidth]{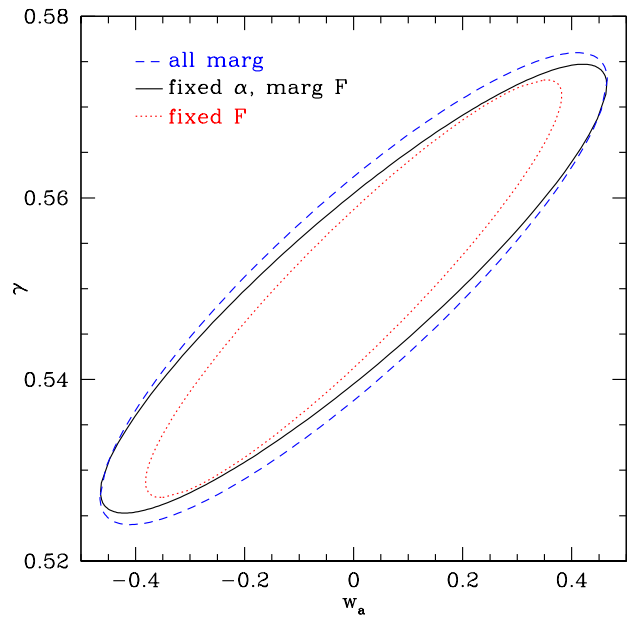}
\caption{Joint 2D $1\sigma$ confidence contours on the dark cosmology 
parameters are 
shown for the cases of all reconstruction parameters being fixed 
(dotted red), marginalized over without redshift dependence as 
in most of the article (solid black), and additionally marginalizing 
over a redshift evolution parameter $\alpha$ (dashed blue).  Here 
$\kmax=0.5\,h$/Mpc.  Note that the fixed $F$ case, shown here centered 
on the true cosmology, could be strongly biased if $F$ was misestimated 
(see Sec.~\ref{sec:bias}). 
}
\label{fig:ellalpha}
\end{figure}

The introduction of redshift dependence through $\alpha$ has little 
impact on the cosmological parameter estimation; the largest correlation 
coefficient of $\alpha$ cosmologically is 0.31, with $\gamma$, and overall 
0.83 with $B_{0.45}$, while $\alpha$ itself is determined to within 0.025.  
Figure~\ref{fig:ellalpha} shows the influence on dark cosmology parameter 
estimation of marginalization over the reconstruction 
parameters with and without redshift dependence, and fixing the 
reconstruction parameters (i.e.\ with a total of 48, 47, or 35 
parameters).  Uncertainties on $w_0$, $w_a$, and $\gamma$ 
increase by only 0.8\%, 0.2\%, 5\% respectively upon including 
$\alpha$.  Other forms of redshift dependence may have different 
cosmological impact, and this deserves further analysis through 
simulations, but the scaling of the characteristic wavenumber as used 
here should give a reasonable first indication.

\subsection{Nonlinear Power Spectrum} \label{sec:halofit} 

The greatest effect of uncertainty in the reconstruction function comes 
from the parameters $A_k$, as seen in Fig.~\ref{fig:fplot} and in 
Sec.~\ref{sec:bias} regarding the parameter bias.  Recall that $A(k)$ 
arises from the nonlinear effects in the density field, and even exists 
when $\mu=0$.  In this limit $A(k)$ acts to map the linear real space 
density power spectrum to the nonlinear regime.  Therefore, if we had a 
robust nonlinear (or quasilinear) real space power spectrum then we would 
have no need of a separate parameter as then $A(k)=1$ (this has been tested 
and found 
accurate to subpercent level by \cite{julithesis}).  Substantial 
effort is going into developing cosmic emulators \cite{emu} that could 
provide accurate nonlinear power spectra, eventually including the 
full set of cosmological parameters and redshifts considered here.  
Since that is still in the future, we consider the nonlinear prescription 
of Halofit \cite{halofit} to get an indication of the potential impact on 
our conclusions. 

The linear power spectrum at a given redshift is fed into Halofit to 
give the approximate nonlinear form.  This removes the need for $A(k)$, 
setting this equal to one for all $k$ and $z$.  As indicated earlier in 
this section, the simulation results from \cite{julithesis} imply that 
for such a normalized $A$ then the quantities $B$ and $C$ become 
substantially redshift independent.  Therefore we do not need any 
hypothetical model such as the $\alpha$ parametrization, making the 
entire analysis more robust.  Furthermore, Halofit includes cosmology 
dependence and so no assumption about universality of $A$ is needed. 

We show the results for cosmological parameter estimation in 
Table~\ref{tab:halofit}, for $\kmax=0.5\,h$/Mpc, for the three cases of 
using the model 
independent approach of fitting for $A(k)$ in bins, using Halofit and $A=1$, 
and using the revised version of Halofit from \cite{12082701} and 
$A=1$.  In all cases we still fit for the binned values of $B$ and $C$. 
The use of functional forms for the nonlinear real space power spectrum 
allows better determination of the cosmological parameters, by 12\%, 
28\%, 12\% for $w_0$, $w_a$, $\gamma$ respectively (15\%, 33\%, 15\% for 
revised Halofit, which has slightly more quasilinear power).  This offers some 
promise for the future use of cosmic emulators, but in this paper we 
prefer to be conservative in the estimations and use the model independent 
approach.

\begin{table}[!htb]
\begin{tabular}{l|cccc}
&$w_0$&$w_a$&$\gamma$&$\Omega_m$\\
\hline 
Fit $A(k)$& $\ $0.0713$\ $ & $\ $0.306$\ $ & $\ $0.0163$\ $& $\ $0.00794$\ $\\ 
Halofit& 0.0624 & 0.220 & 0.0143 & 0.00678\\ 
New Halofit$\ $& 0.0603 & 0.206 & 0.0139 & 0.00608\\  
\end{tabular}
\caption{$1\sigma$ constraints as in Table~\ref{tab:sigkmax}, using 
$\kmax=0.5\,h$/Mpc, but for three different methods of treating 
nonlinearities. 
}
\label{tab:halofit} 
\end{table}

\section{Conclusions} \label{sec:concl} 

With the ability to map galaxy clustering in three dimensions over large 
volumes of the universe comes the necessity for accurate theoretical 
interpretation.  This entails linking the isotropic, linear theory real space 
density power spectrum to the observed anisotropic, nonlinear redshift space 
galaxy power spectrum.  We have investigated here some of the relevant issues 
involving nonlinearities in the density field and velocity effects, using 
the Kwan-Lewis-Linder analytic redshift space reconstruction function 
calibrated from numerical simulations.  

The main question addressed is to what accuracy the anisotropic redshift 
space power spectrum must be known in order to achieve robust cosmological 
conclusions.  We propagate uncertainties in the power spectrum through a 
model independent binning of reconstruction amplitudes with wavenumber and 
assess the effects of deviations from fiducial values.  To avoid 
biasing cosmological parameters such as the dark energy equation 
of state and gravitational growth index requires down to 0.3\% accuracy 
on the reconstruction parameters in the most stringent cases, while smoother 
deviations give more tractable requirements.  Note that it is only those 
deviations that mimic cosmological variations that are most important. 

A more flexible and robust approach is to carry out a global fit for the 
binned reconstruction parameters simultaneously with the cosmological 
parameters, which avoids biasing the results so long as the {\it form\/} 
of the KLL function is a good approximation.  With 8 cosmology parameters 
and 39 systematics parameters we find that a next generation galaxy redshift 
survey such as BigBOSS can tightly and accurately constrain cosmology, for 
example determining the equation of state time variation $w_a$ to 0.3 and 
testing gravity through the growth index $\gamma$ to 3\%.  No external 
priors on the reconstruction parameters are necessary as they are 
selfcalibrated by the survey, most at the subpercent level.  This also 
corresponds to subpercent calibration of the redshift space power spectrum. 

We have tested the robustness of the conclusions by adding redshift 
evolution, which has little effect, varying the number of wavenumber bins, 
and exploring the leverage from increasing the maximum wavenumber used, 
$\kmax$.  Cosmological leverage plateaus by $\kmax=0.4-0.5\,h$/Mpc so the 
KLL form need only apply up to this scale.  We made no assumptions about 
the departure from linearity, allowing the nonlinearity amplitude to float in 
a model independent manner in bins of $k$, but also then analyzed the impact 
of adopting a nonlinear prescription such as Halofit (or its revision).  
This improved the cosmology estimation and offers a promising sign to 
motivate continued development of cosmic emulators for the nonlinear 
power spectrum. 

Several areas exist for further development.  The KLL form has been 
tested for dark matter, but \cite{julithesis} indicates it is successful 
for halos as well.  Eventually this must be extended to galaxies, a major 
undertaking.  On the positive side, we achieved excellent results using 
reconstruction starting from simple linear theory (Kaiser approximation); 
higher order perturbation theory approaches extend the range where 
reconstruction is milder.  Universality, i.e.\ cosmology dependence, of 
the reconstruction is a major topic for future investigation, requiring 
large suites of cosmological simulations, again suited for emulators.  
We have taken a first step toward addressing this effect by exploring 
the influence of using Halofit and new Halofit cosmological dependences for 
the nonlinearity.  Simulations may also enable us to compress the information 
in bins down to a smaller set of parameters.  

Redshift space distortions provide a powerful tool for measuring the 
growth rate of cosmic structure, and delivering insights on the competition 
between the gravitational laws driving clustering and accelerated expansion 
suppressing it.  The results here give encouraging indications, and 
quantitative measures, that theoretical analysis can take into account 
robustly the nonlinear and velocity effects to extract accurate cosmology 
from the forthcoming  large volume redshift surveys.

\acknowledgments 

We thank Sudeep Das, Juliana Kwan, and Alberto Vallinotto for helpful 
discussions.  
This work has been supported by DOE grant DE-SC-0007867 and the Director, 
Office of Science, Office of High Energy Physics, 
of the U.S.\ Department of Energy under Contract No.\ DE-AC02-05CH11231. 
EL acknowledges World Class University grant R32-2009-000-10130-0 through the 
National Research Foundation, Ministry of Education, Science and Technology 
of Korea; JS is supported by the Dark Cosmology Centre, funded by the 
Danish National Research Foundation, and thanks LBNL for additional support.


\end{document}